%% file: graphiq.tex
\documentclass[a4paper,twocolumn,11pt,accepted=2024-08-18]{quantumarticle}
\input{config/config}

\newcommand{\software}{\texttt{GraphiQ}}
\begin{document}
\input{config/title}
\maketitle
\input{text/0abstract}
%\tableofcontents 

\input{text/1introduction}
\input{text/2software_overview}

\input{text/3simulation}
\input{text/4evaluation}
\input{text/5exploration}

\input{text/6conclusion}

\FloatBarrier
\input{text/7back_material}

\FloatBarrier

\appendix
\crefalias{section}{appendix}

\input{text/app_background}
\input{text/app_examples}

\newpage
\bibliographystyle{unsrtnat}
\bibliography{graphiq}

\end{document}

%% file: config/config.tex
\pdfoutput=1
\usepackage{graphicx} % Required for inserting images
\usepackage{amsmath,amssymb,amstext} % Lots of math symbols and environments
\usepackage[utf8]{inputenc}
\usepackage[english]{babel}
\usepackage{url}

\usepackage{breakurl}
\usepackage[colorlinks=true,breaklinks]{hyperref}
\usepackage{listings} % for code blocks
\usepackage[T1]{fontenc} % fix quotation marks in code blocks
\usepackage[frozencache,cachedir=minted-cache, newfloat]{minted}
 \usepackage{float}
\usepackage{caption} % allows some extra formatting for captions
\usepackage{physics}
\usepackage[table]{xcolor}
\usepackage{tabularx,booktabs}
\usepackage{mdframed}
\usepackage[numbers, sort&compress]{natbib}
\usepackage{enumitem}
\usepackage{placeins}
\usepackage{pifont}
\usepackage{datetime}
\newdate{date}{23}{08}{2024}
\usepackage[caption=false,subrefformat=parens,labelformat=simple]{subfig}

\usepackage{algorithm}

\usepackage{algpseudocode}
\makeatletter
\newenvironment{breakablealgorithm}
{% \begin{breakablealgorithm}
	\begin{flushleft}
		\refstepcounter{algorithm}% New algorithm
		\hrule height.8pt depth0pt \kern2pt% \@fs@pre for \@fs@ruled
		\renewcommand{\caption}[2][\relax]{% Make a new \caption
			{\raggedright\textbf{\fname@algorithm~\thealgorithm} ##2\par}%
			\ifx\relax##1\relax % #1 is \relax
			\addcontentsline{loa}{algorithm}{\protect\numberline{\thealgorithm}##2}%
			\else % #1 is not \relax
			\addcontentsline{loa}{algorithm}{\protect\numberline{\thealgorithm}##1}%
			\fi
			\kern2pt\hrule\kern2pt
		}
	}{% \end{breakablealgorithm}
		\kern2pt\hrule\relax% \@fs@post for \@fs@ruled
	\end{flushleft}
}
\makeatother

\usepackage{cleveref}

\crefname{theorem}{Theorem}{Theorems}
\Crefname{theorem}{Theorem}{Theorems}
\crefname{thm}{Theorem}{Theorems}
\Crefname{thm}{Theorem}{Theorems}
\crefname{assump}{Assumption}{Assumptions}
\Crefname{assump}{Assumption}{Assumptions}
\crefname{problem}{Problem}{Problems}
\Crefname{problem}{Problem}{Problems}
\crefname{conjecture}{Conjecture}{Conjectures}
\Crefname{conjecture}{Conjecture}{Conjectures}
\crefname{proposition}{Proposition}{Propositions}
\Crefname{proposition}{Proposition}{Propositions}
\crefname{prop}{Proposition}{Propositions}
\Crefname{prop}{Proposition}{Propositions}
\crefname{cor}{Corollary}{Corollaries}
\Crefname{cor}{Corollary}{Corollaries}
\crefname{lem}{Lemma}{Lemmas}
\Crefname{lem}{Lemma}{Lemmas}
\crefname{definition}{Definition}{Definitions}
\Crefname{definition}{Definition}{Definitions}
\crefname{defn}{definition}{definitions}
\Crefname{defn}{Definition}{Definitions}
\crefname{remark}{Remark}{Remarks}
\Crefname{remark}{Remark}{Remarks}
\crefname{rmk}{Remark}{Remarks}
\Crefname{rmk}{Remark}{Remarks}
\crefname{example}{Example}{Examples}
\Crefname{example}{Example}{Examples}
\crefname{table}{Table}{Tables}
\Crefname{table}{Table}{Tables}
\crefname{align}{}{}
\Crefname{align}{}{}
\crefname{listing}{codeblock}{codeblocks}
\Crefname{listing}{Codeblock}{Codeblocks}
\crefname{equation}{eq.}{eqs.}
\Crefname{equation}{Eq.}{Eqs.}
\crefname{step}{Step}{Steps}
\Crefname{step}{Step}{Steps}
\crefname{protocol}{protocol}{protocols}
\Crefname{protocol}{Protocol}{Protocols}
\crefname{algorithm}{algorithm}{algorithms}
\Crefname{algorithm}{Algorithm}{Algorithms}

\newenvironment{aeq}{\begin{equation}
	\begin{aligned}
}{
	\end{aligned}
\end{equation}}

\lstdefinestyle{codeblock}{
  backgroundcolor=\color{backcolour},   
  commentstyle=\color{codegreen},
  keywordstyle=\color{blue},
  numberstyle=\tiny\color{codegray},
  stringstyle=\color{codepurple},
  basicstyle=\footnotesize,
  escapechar=\¢,escapebegin=\color{purple}, % highlighting for decorator
  otherkeywords={with},
  breakatwhitespace=false,         
  breaklines=true,                 
  captionpos=b,                    
  keepspaces=true,
  language=Python,
  numbers=right,                    
  numbersep=5pt,                  
  showspaces=false,                
  showstringspaces=false,
  showtabs=false,                  
  tabsize=2,
  basicstyle=\ttfamily\footnotesize
}
\newcommand{\classname}[1]{\texttt{#1}}

\usepackage{tabularx}

\usepackage[format=plain,labelfont={bf,it},textfont=it]{caption}
\newenvironment{code}{\captionsetup{type=listing}}{}
\SetupFloatingEnvironment{listing}{name=Codeblock}
\newmintedfile[pythonfile]{python}{
	bgcolor=gray!6,
	fontfamily=tt,
	gobble=0,
	fontsize=\small,
	breaklines=true,
    frame=leftline,
	framerule=0.4pt,
	funcnamehighlighting=true,
	tabsize=2,
	obeytabs=false,
	mathescape=false
	samepage=false, %with this setting you can force the list to appear on the same page
	showspaces=false,
	showtabs =false,
	texcl=false,
	linenos=true,
	xleftmargin=1pt,
	numbersep=1pt
}

\definecolor{tablehead}{HTML}{CEF6F5}
\definecolor{xgreen}{HTML}{119A68}

%% file: config/title.tex
\title{GraphiQ: Quantum circuit design for photonic graph states}

\author{Jie Lin$^{*}$}
\orcid{0000-0003-1750-2659}
\affiliation{Quantum Bridge Technologies Inc., 108 College St., Toronto, ON, Canada}
\affiliation{Department of Electrical and Computer Engineering, University of Toronto, 10 King’s College Road, Toronto, ON, Canada}
\author{Benjamin MacLellan$^{*}$}
\orcid{0000-0001-7576-8020}
\affiliation{University of Waterloo, Department of Physics \& Astronomy, 200 University Ave., Waterloo, ON, Canada}
\affiliation{Institute for Quantum Computing, 200 University Ave., Waterloo, ON, Canada}
\affiliation{Ki3 Photonics Technologies, 2547 Rue Sicard, Montreal, QC, Canada}
\author{Sobhan Ghanbari$^{*}$}
\orcid{0009-0004-0318-4914}
\affiliation{Quantum Bridge Technologies Inc., 108 College St., Toronto, ON, Canada}
\affiliation{Department of Physics, University of Toronto, 60 St George St., Toronto, ON, Canada}
\author{Julie Belleville}
\affiliation{Ki3 Photonics Technologies, 2547 Rue Sicard, Montreal, QC, Canada}
\author{Khuong Tran}
\affiliation{Ki3 Photonics Technologies, 2547 Rue Sicard, Montreal, QC, Canada}
\author{Luc Robichaud}
\orcid{0000-0001-8974-0504}
\affiliation{Quantum Bridge Technologies Inc., 108 College St., Toronto, ON, Canada}
\affiliation{Department of Electrical and Computer Engineering, University of Toronto, 10 King’s College Road, Toronto, ON, Canada}
\author{Roger G. Melko}
\orcid{0000-0002-5505-8176}
\affiliation{University of Waterloo, Department of Physics \& Astronomy, 200 University Ave., Waterloo, ON, Canada}
\affiliation{Perimeter Institute for Theoretical Physics, 31 Caroline St N., Waterloo, ON, Canada}
\author{Hoi-Kwong Lo}
\orcid{0000-0002-0340-4989}
\affiliation{Quantum Bridge Technologies Inc., 108 College St., Toronto, ON, Canada}
\affiliation{Department of Electrical and Computer Engineering, University of Toronto, 10 King’s College Road, Toronto, ON, Canada}
\author{Piotr Roztocki}
\orcid{0000-0002-1068-6355}
\affiliation{Ki3 Photonics Technologies, 2547 Rue Sicard, Montreal, QC, Canada}
	\def\thefootnote{*}\footnotetext{These authors contributed equally to this work}\def\thefootnote{\arabic{footnote}}
\date{\displaydate{date}}

%% file: text/0abstract.tex
\begin{abstract}
    \software{} is a versatile open-source framework for designing  photonic graph state generation schemes, with a particular emphasis on photon-emitter hybrid circuits. Built in Python, \software{} consists of a suite of design tools, including multiple simulation backends and optimization methods. The library supports scheme optimization in the presence of circuit imperfections, as well as user-defined optimization goals. Our framework thus represents a valuable tool for the development of practical schemes adhering to experimentally-relevant constraints. As graph states are a key resource for measurement-based quantum computing, all-photonic quantum repeaters, and robust quantum metrology, among others, we envision \software{}'s broad impact for advancing quantum technologies. 
\end{abstract}

%% file: text/1introduction.tex
\section{Introduction}\label{sec:intro}

\begin{figure*}[t]	
    \centering
    \includegraphics[width=\linewidth]{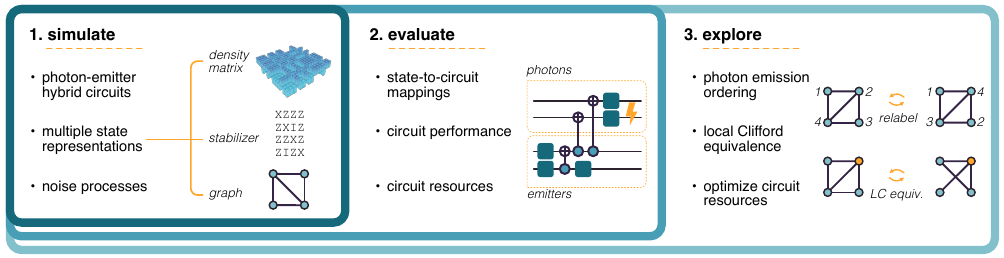}
	\caption{\software{} software framework and its three core use cases: the simulation, evaluation, and exploration of quantum circuits that generate photonic graph states.}	\label{fig:architecture}
\end{figure*}

% Graph states are important resources in QIP
\software{} is an open-source software framework for the design of photonic quantum circuits, with a particular focus on the realistic generation of entangled graph states in hybrid photon-emitter platforms. Graph states, a family of entangled quantum states, are a key resource in a variety of quantum information applications, including entanglement distribution and quantum networking \cite{azuma_all-photonic_2015, hilaire_resource_2021}, quantum error correction \cite{looi_quantum-error-correcting_2008,bell_experimental_2014}, measurement-based quantum computing \cite{raussendorf_one-way_2001,raussendorf_measurement_2003,varnava_loss_2006}, fusion-based quantum computing \cite{Nielsen_fusion_2004,Segovia_fusion_2015,bartolucci_fusion_2021}, and quantum metrology \cite{shettell_graph_2020}.
Among the various platforms leveraged for quantum technologies, photonics has unique advantages towards graph state generation, such as qubits with long coherence times, compatibility with well-developed integrated and fiber optic infrastructures, and the ability to be distributed spatially (e.g., between different network nodes).

%probabilistic vs deterministic methods
Generating photonic graph states can be realized with two main approaches: probabilistic and deterministic. In the probabilistic case, fusion gates \cite{browne_resource-efficient_2005, Ewert2014}, implemented with linear optics, detectors and post-selection, are used to build up graph states from small entangled states. However, given the probabilistic nature of the fusion, the resources required scale exponentially with the graph state size \cite{Pant_ProbScaling_2017}. On the other hand, deterministic approaches leverage entanglement operations between emitters, such as quantum dots, trapped ions or nitrogen-vacancy centers in diamond \cite{Russo_QD2Dgraphs_2019}, for direct graph state generation without the need for probabilistic fusion. Recent experimental demonstrations using such architectures \cite{lindner_proposal_2009} have reached impressive milestones, e.g., the generation of 10-qubit linear cluster states \cite{schwartz_deterministic_2016,cogan_deterministic_2021} and 14-qubit Greenberger–Horne–Zeilinger (GHZ) states \cite{thomas_efficient_2022}. There are also various theoretical proposals on quantum-emitter-based methods for the generation of 2-dimensional graph states  \cite{Economou_QD2Dgraphs_2010,Buterakos_QD2Dgraphs_2017,Segovia_QD2Dgraphs_2019,Russo_QD2Dgraphs_2019}. Due to limitations on the coherence time of quantum emitters and challenges in their coupling, graph states generated by such approaches are still too small to be useful for many practical applications. Searching for resource-efficient generation schemes obeying realistic experimental constraints is thus an active goal for the research community.

% Photonics as a platform for quantum technologies has many benefits, but suffers from a lack of photon-photon interaction.
Nevertheless, designing the quantum circuit(s) appropriate for the generation of a particular graph state is a non-trivial task, especially as the size of the graph state grows. 
This is in part due to the lack of usable photon-photon interactions; using linear optics and post-selection to perform entangling operations, after photon emission, is impractical due to the rapid scaling in the number of required components and the drastic decrease in success probability. 
Rather, entanglement can be created through interactions between emitter qubits, and subsequently transferred to photonic qubits through the emission process (represented as a CNOT gate between emitter and photon) \cite{Russo_QD2Dgraphs_2019}. 
However, identifying the required sequence of quantum operations, for both the emitter and photonic qubits, to realize a particular entangled state is challenging. 

In recent years, a rich ecosystem of software packages for simulating and designing quantum circuits has been developed \cite{pennylane_v4_2018, gray_quimb_2018, Killoran_strawberryfields_2019, luo_yao_2020, qiskit2024, heurtel_perceval_2023}. Such open-source toolboxes have been central to progress in quantum information science, allowing a broad research community to use and collaborate using a common set of tool-chains. While these toolboxes may have similar basic components for simulating quantum circuits and quantum states, they typically focus on specific qubit architectures or computation models. For example, PennyLane \cite{pennylane_v4_2018} is for differential programming of quantum computers and its core feature is to compute gradients of variational quantum circuits; Quimb \cite{gray_quimb_2018} is designed for quantum information and many-body calculations; Strawberry Fields \cite{Killoran_strawberryfields_2019} focuses on continuous-variable systems; Perceval \cite{heurtel_perceval_2023} is for discrete-variable photonic quantum computing, focusing on quantum circuit simulation. No packages thus far have been tailored to designing and optimizing hybrid photonic quantum circuits for photonic graph state generation. Very recently, there is some recent progress in optimizing protocols for graph state generation \cite{li_photonic_2022, Lee2023}. Ref. \cite{Lee2023} introduces \texttt{OptGraphState}, a Python software to study fusion-based graph state generation. In terms of deterministic approaches, Ref. \cite{li_photonic_2022} uncovers a method for constructing circuits that minimize the number of required emitter qubits. It should be noted that further optimizations are still possible to address different experimental challenges. Nevertheless, the task of designing optimized photon-emitter circuits remains challenging, especially if the constraints of real experimental devices (e.g., photon loss, noise) are to be considered. In this paper, we focus on the deterministic approaches.

Here, we introduce \software\footnote{This paper refers to \software{} version 0.1.0.}, an open-source toolbox for hybrid photon-emitter schemes, implemented in Python, particularly suited for the study and design of practical graph state generation circuits. The framework (\Cref{fig:architecture}) comprises a suite of tools towards this end, centered on three main use-cases:  
\begin{enumerate}
    \item[$\bullet$] \textbf{Simulation} of photon-emitter quantum circuits, with multiple backend representations for quantum states (density matrix, stabilizer, and graph);
    \item[$\bullet$] \textbf{Evaluation} of circuit performance and required resources, alongside benchmarking tools;
    \item[$\bullet$] \textbf{Exploration} and optimization of photonic circuits, governed by experimental constraints, user-defined metrics, and target applications.
\end{enumerate} 

Our use of the framework for investigating circuit optimization via local operations \cite{Ghanbari2024} has yielded significant improvements in generation scheme designs for the photonic-emitter platform. The package is, moreover, readily adaptable to new hardware platforms, applications, and optimization methods. 

%Paper outline
The remainder of this paper is outlined below. In \Cref{sec:overview}, we summarize the architecture of \software{} and its key features. We then discuss the circuit simulation module of \software{} in \Cref{sec:simulation}. In \Cref{sec:evaluation}, we discuss evaluating the performance of a given quantum circuit using various circuit- and graph-based metrics. In \Cref{sec:exploration}, we present optimization schemes currently implemented in \software. Lastly, we provide concluding remarks in \Cref{sec:conclusion}. Additional background and implementation details are provided in \Cref{app_sec:background} for reference. We include examples accomplished using \software{} in \Cref{app_sec:examples}.

%% file: text/2software_overview.tex
\section{Overview of \software}\label{sec:overview}

\begin{figure}[t]
%\noindent\fbox{\parbox{0.95\linewidth}{
\begin{mdframed}
What can \software{} do?
\begin{enumerate}[leftmargin=0.5cm]
\item[$\bullet$] Simulate the output of noisy quantum circuits, with emitted quantum states represented as density matrices, stabilizer tableaux, or graphs. See \Cref{sec:simulation}.
\item[$\bullet$] Convert quantum states from one representation to another.
\item[$\bullet$] Find quantum circuits that produce a target graph state and optimize them with respect to custom state/circuit performance metrics. A set of physical rules and experimental constraints can be imposed on the resultant circuits. See \Cref{sec:exploration}.
\item[$\bullet$] Find alternative circuits for the generation of the same target graph state and/or the graphs isomorphic to it.
\item[$\bullet$] Check for the local Clifford equivalency of graph states.
\item[$\bullet$] Find the local Clifford equivalency class (LC orbit) for a given graph.
\item[$\bullet$] Evaluate state, graph theoretical, and circuit metrics.
\item[$\bullet$] Visualize quantum states and circuits.  
\end{enumerate}
\end{mdframed}
\end{figure}

\begin{figure}[H]
    \centering
\includegraphics[width=0.9\linewidth]{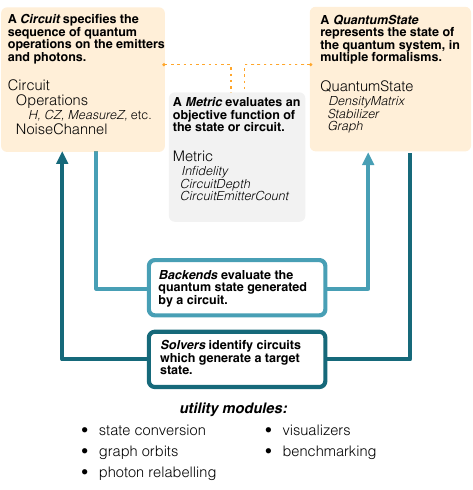}
	\caption{\software{} interface overview. It consists of several modules including state/circuit representation, noise simulation, metric evaluation and various utility modules in addition to optimization solvers.}
	\label{fig:interface}
\end{figure}

\Cref{fig:interface} gives an overview of the package interface and highlights its main classes and modules. The development of \software{} is focused on four design pillars:
\begin{enumerate}
\item[$\bullet$]\textbf{Versatility:} 
    Support for a broad range of tasks related to circuit design, spanning simulation through to visualization and benchmarking;
\item[$\bullet$] \textbf{Accuracy:} 
    Experimental realism is upheld by an extensive library of circuit noise models, as well as by imposing hardware constraints on designed circuits;
\item[$\bullet$] \textbf{Modularity:} Module independence enables fast tailoring of software for custom use cases;
\item[$\bullet$] \textbf{Extensibility:} Users can rapidly prototype and include new, custom optimization algorithms, physical circuit constraints, and state/circuit metrics. 
\end{enumerate}

Tutorials and examples of \software{} are available on \url{https://github.com/graphiq-dev/graphiq}. More information about the full documentation can be found in \url{https://graphiq.readthedocs.io/en/latest/}.

%% file: text/3simulation.tex
\section{Simulation of quantum circuits}\label{sec:simulation}

Simulating the quantum state produced by a given circuit is prerequisite to its optimization. \software{}'s simulation modules comprise representations of quantum states, circuits, and noise.

\subsection{States}
The underlying quantum state of emitters and photons, represented by the evolution of an initial state through the circuit, can be modelled with different classical simulation techniques. \software{} supports multiple state representations, providing flexibility for different use cases. The state representation module comes with a wrapper class \classname{QuantumState} that acts as a common application programming interface (API) for users. 
Currently it supports several backends: \textbf{density matrix}, \textbf{stabilizer}, \textbf{mixed stabilizer}, or \textbf{graph}. \Cref{code:quantum_state} shows one way to initialize \classname{QuantumState} objects for initial data in the form of a \classname{Graph} object from the \classname{NetworkX} package. 
\begin{code}
	\pythonfile{code/states.py}
	\caption{Using a \classname{NetworkX} graph to initialize two \classname{QuantumState} objects: one with the density matrix backend and the other with the graph backend.}
 \label{code:quantum_state}
\end{code}

We discuss briefly each representation, their use cases, and limitations below.

A quantum state in a finite-dimensional Hilbert space can be represented as a positive semi-definite matrix with unit trace. The \textbf{density matrix} $\rho$ of an $n$-qubit state is an element of $\mathbb{C}^{2^n \times 2^n}$ (see Ref. \cite[Section 2.4]{nielsen_quantum_2010} for review).
The density matrix formalism captures all information about the underlying quantum state and enables the state to evolve under arbitrary unitary operations, quantum measurements, and general noise processes. However, due to the exponential scaling of the size of the density matrix with the number of qubits, it becomes an impractical representation to describe large states as both the memory storage and running time for state simulations scale as O($2^n$). Consequently, the density matrix backend is only suitable for states with a small number of qubits (e.g., $n \leq 10$). 

In turn, the \textbf{stabilizer} formalism \cite{gottesman_stabilizer_1997,aaronson_improved_2004} enables the efficient representation of so-called stabilizer states, a subset of quantum states. As graph states form a subset of stabilizer states, this representation is suitable for their study. An $n$-qubit stabilizer state is uniquely defined by its stabilizer group which has $n$ independent generators. More background information about the stabilizer formalism can be found in \Cref{app_subsec:stabilizer}. See also Refs. \cite{aaronson_improved_2004,audenaert_entanglement_2005}. 
In particular, \software{} adopts the formalism in Ref. \cite{aaronson_improved_2004}, keeping in memory the stabilizer and destabilizer generators\footnote{As defined in \cite{aaronson_improved_2004}, destabilizer generators are Pauli operators such that destabilizer and stabilizer generators together generate the entire $n$-qubit Pauli group up to four global phases $\pm 1, \pm i$.} of the $n$-qubit stabilizer state, as an $2n \times 2n$ binary matrix, which scales much better relative to density matrices. For each generator, \software{} also uses 2 bits to keep track of four possible phases. We note that generators corresponding to the stabilizer part can take only $\pm~1$ phases, while the destabilizer generators can take on both $\pm~1$ and $\pm~i$ phases.

Graph states are, by construction, defined by a \textbf{graph} where the nodes represent qubits and the connecting edges represent entanglement between the qubits.
In this representation, graph states correspond to simple, undirected graphs, which can be stored in memory as adjacency matrices, or node and edge sets. As the size of node-edge sets grows with the connectivity of the graph while adjacency matrices remain constant in size for a given qubit number, node-edge sets are more favorable for graphs with low connectivity, while adjacency matrices are more favorable for high connectivity cases. 

Given that the graph representation is very visual, it is useful for understanding the entanglement structure of a given state, which is not always so obvious in the case of the density matrix or stabilizer representations. A graph representation also makes it more intuitive to look at graph actions like local complementation. The action of local complementation on a graph corresponds to local Clifford operations on the corresponding graph state (see \Cref{app_sec:background}). 

Since graph states form a proper subset of stabilizer states, to empower this representation to represent all stabilizer states, it is necessary to amend this representation. As any stabilizer state can be converted to a graph state by local Clifford operations \cite{van_den_nest_graphical_2004}, \software{} accomplishes this task by including a record of required local Clifford operations in each node of the graph.

It is worth noting that the stabilizer and graph representations discussed so far can represent only pure states. However, when noise is present, it is usually necessary to consider \textbf{mixed states}. Unlike the density matrix representation, which can naturally represent mixed states, the stabilizer and graph representations must be amended to support mixed states. A simple approach, implemented in \software{}, is to keep track of an ensemble of pure states that can form the mixed state of interest.

\software{} also offers functions to convert different state representations. More information can be found in \Cref{app_subsec:state_conversion}.

\subsection{Circuits}

\begin{figure}[t]
\centering
\subfloat[]{\label{fig:graph_linear3}\includegraphics[width=0.35\linewidth]{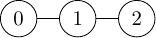}}\\
\subfloat[]{\label{fig:circuit_linear3}\includegraphics[width=0.95\linewidth]{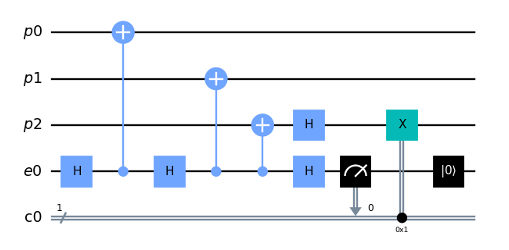}}
	\caption{(a) Graph representation for a 3-qubit linear cluster, where each node represents a qubit initialized in a $\ket{+}$ state, and each edge represents a CZ gate. Each node is assigned a unique ID. (b) A quantum circuit that generates the above cluster state by using one emitter. p0-p2 represent 3 photonic qubits corresponding to those in (a), e0 represents the emitter qubit, and c0 represents a classical bit.}
\label{fig:simulation_example}
\end{figure}

Quantum circuits describe a sequence of operations applied to qubits. \software{} is developed with a particular focus on emitter-photonic hybrid circuits, meaning that circuits are composed of three register blocks: emitter qubits, photonic qubits, and classical bits. For example, \Cref{fig:circuit_linear3} illustrates an emitter-photonic circuit that produces the 3-qubit linear cluster state shown in \Cref{fig:graph_linear3}. In \software, the \classname{CircuitDAG} class can be used to store a quantum circuit (internally represented as a directed acyclic graph, or DAG), enabling easy modification. \Cref{code:circuit} demonstrates how quantum gates can be added into a circuit one by one. See \Cref{app_sec:code_blocks} for further code examples, including circuit simulation. 
\begin{code}
	\pythonfile{code/circuit.py}
	\caption{Construction of a quantum circuit by adding gates.}
 \label{code:circuit}
\end{code}

Additionally, users can also insert, remove, or replace a quantum gate at a specified location. In addition to direct manipulation of quantum circuits, it is also possible to import a quantum circuit from Open Quantum Assembly Language (OpenQASM) files \cite{cross_open_2017}. Likewise, \software{} circuit objects can be serialized to OpenQASM files for export to other toolchains, such as external visualizers and simulators \cite{qiskit2024}. 

\subsection{Noise simulation}
In practice, noise and experimental imperfections affect all quantum information tasks. The capacity to simulate noise is prerequisite to modeling schemes resembling true experimental conditions.

\software{} is flexible in its modeling, allowing users to simulate noise within and/or between quantum gates. This respectively corresponds to replacing circuit gates with their noisy versions, or to inserting noise before/after quantum gates (i.e., noise between operations, like during photon propagation or storage). 

The software presently supports noise types common and relevant to photonic graph state generation, e.g., depolarizing noise and photon loss.
\Cref{code:noise} illustrates an example of applying photon loss and depolarizing noise to a state. 
\begin{code}
	\pythonfile{code/noise.py}
 \caption{Applying photon loss with a loss probability of $0.1$ to the first qubit of a 2-qubit state and then applying a depolarizing channel with a depolarizing probability of $0.02$ to the same qubit.}
 \label{code:noise}
\end{code}

Moreover, the included noise models can be readily extended and customized to account for user-specific hardware considerations. Extended discussion and specific examples on noise simulation in \software{} can be found in \Cref{app_subsec:noise_models}.

%% file: text/4evaluation.tex
\section{Evaluation of circuits}\label{sec:evaluation}

When optimizing a quantum circuit for the generation of a target graph state, it is important to be able to quantify the circuit performance using experimentally relevant figures of merit, which are often specific to the user hardware. 
 
\software{} enables great flexibility in evaluating circuits with figures of merit (defined by the user) that can be based on the output quantum state and/or the circuit itself. The software also allows defining a weighted average of a list of metrics or custom functions. Below, we discuss already-implemented metrics based on quantum circuits and output states, as well as the algorithms that leverage them to converge on performant generation schemes for targeted graph states (i.e., state-to-circuit mappings).

\subsection{Circuit metrics}\label{sec:circuit_metric}
Reducing the total number of gates or the longest path in a circuit (i.e., the circuit depth) is of general interest as it corresponds to a reduction in experimental complexity and cost. However, one may care about reducing specific gate types due to the experimental difficulty of implementing them. For example, in the context of generating photonic graph states using solid-state emitters, operations on emitter qubits are typically more challenging to realize in practice. Photonic-emitter circuit designers may thus want to minimize the circuit depth on emitter qubits, the number of CNOT between emitter qubits, as well as the number of emitter qubits, among others. 

\software{} thus supports circuit metrics such as:
\begin{enumerate}
    \item[$\bullet$] Total number of unitaries,
    \item [$\bullet$] Number of emitter qubits,
    \item[$\bullet$] Total number of emitter-emitter CNOT gates,
    \item[$\bullet$] Circuit depth: the length of the longest path in the quantum circuit,
    \item[$\bullet$] Emitter depth: the maximum number of gates applied on an emitter in the circuit between two consecutive resets of that emitter,
    \item[$\bullet$] Emitter history: the maximum number of gates applied on a single emitter.
\end{enumerate}
The implementation of custom user-defined metrics of the generation circuit is supported in \software{} and straightforward.

\Cref{code:circuit_metric} shows an example of evaluating a metric, which has a common interface for both circuit and state metrics. 
\begin{code}
	\pythonfile{code/circuit_metric.py}
 \caption{Creating and evaluating the maximum emitter depth metric on a circuit.}
 \label{code:circuit_metric}
\end{code}

\subsection{State metrics}\label{sec:state_metric}
Fidelity quantifies how alike two quantum states are to one another. Thus, for the task of designing circuits that output a specific target quantum state, circuit performance can effectively be quantified by comparing the actual versus targeted output state via fidelity.

For two states $\rho_{o}$ and $\rho_{t}$, corresponding here to the output and target states respectively, the fidelity is defined as 
\begin{aeq}
 F(\rho_{o}, \rho_{t}) = \tr(\sqrt{\sqrt{\rho_{o}}\rho_{t}\sqrt{\rho_{o}}})^2. 
\end{aeq}
As it is natural to minimize a cost function in an optimization algorithm rather than maximizing, \software{} implements infidelity as one of its state metrics, defined as $1-F(\rho_{o}, \rho_{t})$. 
While calculating fidelity is straightforward in the density matrix representation, it is more complicated to do so in the stabilizer representation. Here, \software{} leverages an algorithm based on the one presented in Ref. \cite{garcia_efficient_2012}. 

Other state metrics such as the trace distance are also included in \software{}.

\subsection{State-to-circuit mapping}\label{sec:state_to_circuit}

 In \software{}, a solver is an algorithm that maps a user-defined graph state to a circuit that can generate it, with its performance assessed by a cost function also chosen by the user. The cost function can be composed of state and circuit metrics, e.g., maximizing the output graph state fidelity or minimizing the number of CNOT gates in the circuit. Additionally, each solver accounts for physical/platform constraints to guarantee that any output quantum circuit is realizable. For photonic-emitter circuits, the main constraints relate to the lack of photon-photon interactions; entanglement between photons is generated indirectly by entangling emitter qubits and emitting the photons, creating photon-emitter entanglement. This means that any two-qubit gates (except CNOT gates used to model photon emission) are only allowed between two \textit{emitter} qubits. 

To use a solver, a user specifies the target graph state, cost function, as well as the preferred state representation/compiler. The solver outputs a list of potential quantum circuits, ranked by cost function value. In other words, each solver gives a state-to-circuit mapping for a given choice of cost function. The quality of the mapping will depend on the choice of solver and cost function. From a given state-to-circuit mapping, additional techniques can be applied to derive more advanced solvers; we discuss these further in \Cref{sec:exploration} and demonstrate an application in Ref. \cite{Ghanbari2024}.

The current version of \software{} has multiple solvers implemented. Solvers can loosely be grouped into probabilistic, deterministic, or hybrid approaches. Probabilistic approaches are based on random perturbations to successive generations of quantum circuits, e.g., \classname{EvolutionarySolver}, which leverages evolutionary algorithms. In turn, deterministic approaches execute rule sets without randomness. An example of this is \classname{TimeReversedSolver}, which deterministically finds a quantum circuit for a target graph state. This solver is based on the method of Li et al. presented in Ref. \cite{li_photonic_2022}. It uses time-reversed processes to fully disentangle the target graph state with the aid of so-called height function to guide the optimization. The resulting quantum circuit is obtained by the inverse ordering of gates found in this process. Noteworthily, for a given photon emission ordering, it outputs a single unique circuit which uses the minimum number of quantum emitters required to generate said graph state (making it a great basis for further exploration/optimization, see \Cref{sec:exploration}). 
Finally, hybrid approaches use a combination of deterministic and probabilistic algorithms, e.g., using deterministic solver outputs as a starting point for random perturbations. 

In terms of probabilistic solvers, \software{} currently has  \classname{RandomSearchSolver} and \classname{EvolutionarySolver}, which are suitable for exploring graph states of small size. They are based on applying random operations (which includes addition, removal, or replacement of a quantum gate) to a population of candidate circuits, evaluating the relevant metrics to assign a score to each circuit, and selecting circuits with the highest scores in each iteration. Deterministic solvers include \classname{TimeReversedSolver}, which is the implementation of Ref. \cite{li_photonic_2022}'s algorithm that allows the use of minimal number of emitters. \classname{HybridEvolutionarySolver} is an example of a hybrid solver, which uses the resulting circuit from a deterministic solver (e.g. \classname{TimeReversedSolver}) as a seed, and then applies the evolutionary algorithm similar to \classname{EvolutionarySolver}. Finally, there is \classname{AlternateTargetSolver} that we will discuss in \Cref{sec:exploration}. More details about each solver can be found in the documentation. Community contributions are welcome in \software{} to centralize more sophisticated solver algorithms as they develop.

%% file: text/5exploration.tex
\section{Exploration and optimization of quantum circuits}\label{sec:exploration}

The circuits that generate a target graph state are generally not unique, with many potentially suited for the task to varying degree. Users often aim to simultaneously satisfy several objectives within a design: experimental platform constraints, performance for a specific application, noise robustness, cost of the realization, among others. A central goal of \software{} is enabling exploration across the space of candidate circuits towards practical and optimized designs. The framework includes several tools to support this search including modules for photon emission reordering and local Clifford equivalency, as well as solvers that integrate such modules into optimization workflows. 

\subsection{Photon emission ordering}
If the photons of a graph can be considered indistinguishable, it becomes possible to consider different ways to label the photons in the graph. The nodes are numerically labeled, corresponding to the order of photon emission, with different photon emission orderings naturally corresponding to different quantum circuits that produce states with the same entanglement structure \cite{li_photonic_2022}. For applications that are not sensitive to photon emission ordering, it is thus possible to utilize this degree of freedom for exploration and optimization. 

\software{} provides users with the option to automatically consider emission reordering in the search for suitable circuits, generally leading to more circuit candidates for users to rank and choose from. Further details on the emission ordering module can be found in \Cref{app_subsec:emission_ordering}.

\subsection{Local Clifford equivalency}
Local Clifford gates are tensor products of one-qubit Clifford gates and are typically easy and inexpensive to apply experimentally. Interestingly, applying local Clifford gates to a graph state is closely related to local complementation actions on the corresponding graph \cite{van_den_nest_graphical_2004}. By applying local Clifford operations onto it, a target graph state can be made to span a larger space of states. These additional target states differ, but are locally equivalent to the original target graph state via local Clifford operations. As these can be efficiently transformed into the original target state through single-qubit gates on the photons, the overhead to consider the alternative graphs is minimal. However, using local Clifford equivalent graph states as targets may lead some solvers to produce alternative quantum circuits, useful for optimization. 

\software{} includes modules for exploring local Clifford equivalency to complement the search for quantum circuits, also appending the necessary local Clifford gates. Further details on the module can be found in \Cref{app_subsec:lc}. 

\subsection{Optimization workflows}
\begin{figure*}[t]
    \centering
    \includegraphics[width=\linewidth]{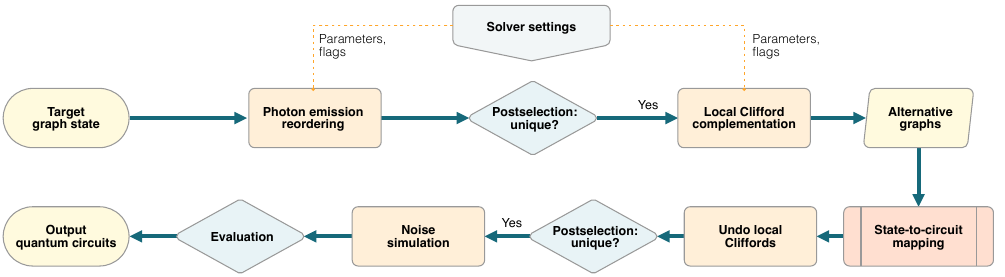}
	\caption{Workflow of \classname{AlternateTargetSolver}. After a target state is provided, photon emission ordering and local Clifford equivalency modules are used to create alternative graph targets. A state-to-circuit mapping algorithm is used to then obtain a set of candidate quantum circuits. These circuits are post-selected to remove redundancy, simulated with user-specified noise sources, and evaluated in performance by user-defined metrics to identify the best circuits.}
 \label{fig:pipeline}
 \end{figure*}

\software{} supports user-defined optimization workflows built up of the solvers and tools described in the preceding sections, as well as those added by users. In particular, as discussed in \Cref{sec:state_to_circuit}, existing solvers can be used as a building block for more advanced exploration and optimization algorithms. 

As an example, we describe \classname{AlternateTargetSolver}, available in \software{}, which leverages photon emission ordering, local Clifford equivalency, and deterministic state-to-circuit mapping to generate an array of circuit candidates. Users can then rank these candidate circuits by their preferred metric and select the best one to use. The workflow of this solver is shown in \Cref{fig:pipeline}.
First, emission reordering and local Clifford complementation are used to create a list of alternatives to the initial target graph state. A deterministic solver like \classname{TimeReversedSolver} is then used as a state-to-circuit mapping to run a noise-free optimization, outputting a set of alternative circuits. The local Clifford gates necessary to complete the quantum circuits (to produce the original target state) are then found. A post-selection procedure eliminates redundant circuits from the set. Finally, circuit noise is simulated prior to a final evaluation of the circuits' performance metrics and ranking. An example based on \classname{AlternateTargetSolver} is showcased in \Cref{app_subsec:example_alternate_graph}, demonstrating its ability to produce alternative target graph generating circuits. A more detailed study can also be found in \cite{Ghanbari2024}.

The tools in \software{} are well suited to user-customized workflows allowing, as in the case of \classname{AlternateTargetSolver}, to explore and converge on experimentally-practical circuit candidates.

\subsection{Runtime}

\begin{figure}[H]
	\centering
	\subfloat[Linear cluster state]{
		\includegraphics[width=\linewidth]{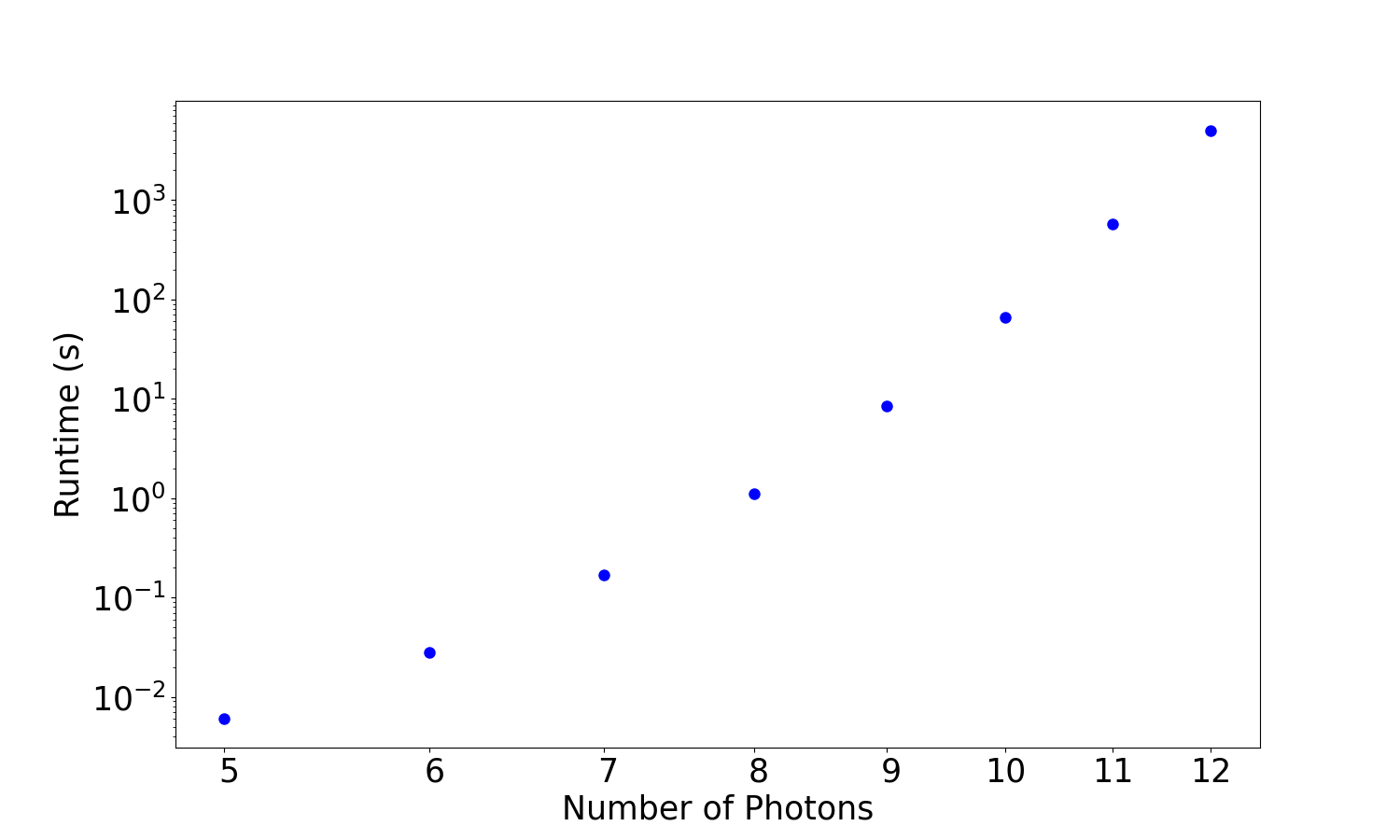}}\\
	\subfloat[Repeater graph state]{\includegraphics[width=\linewidth]{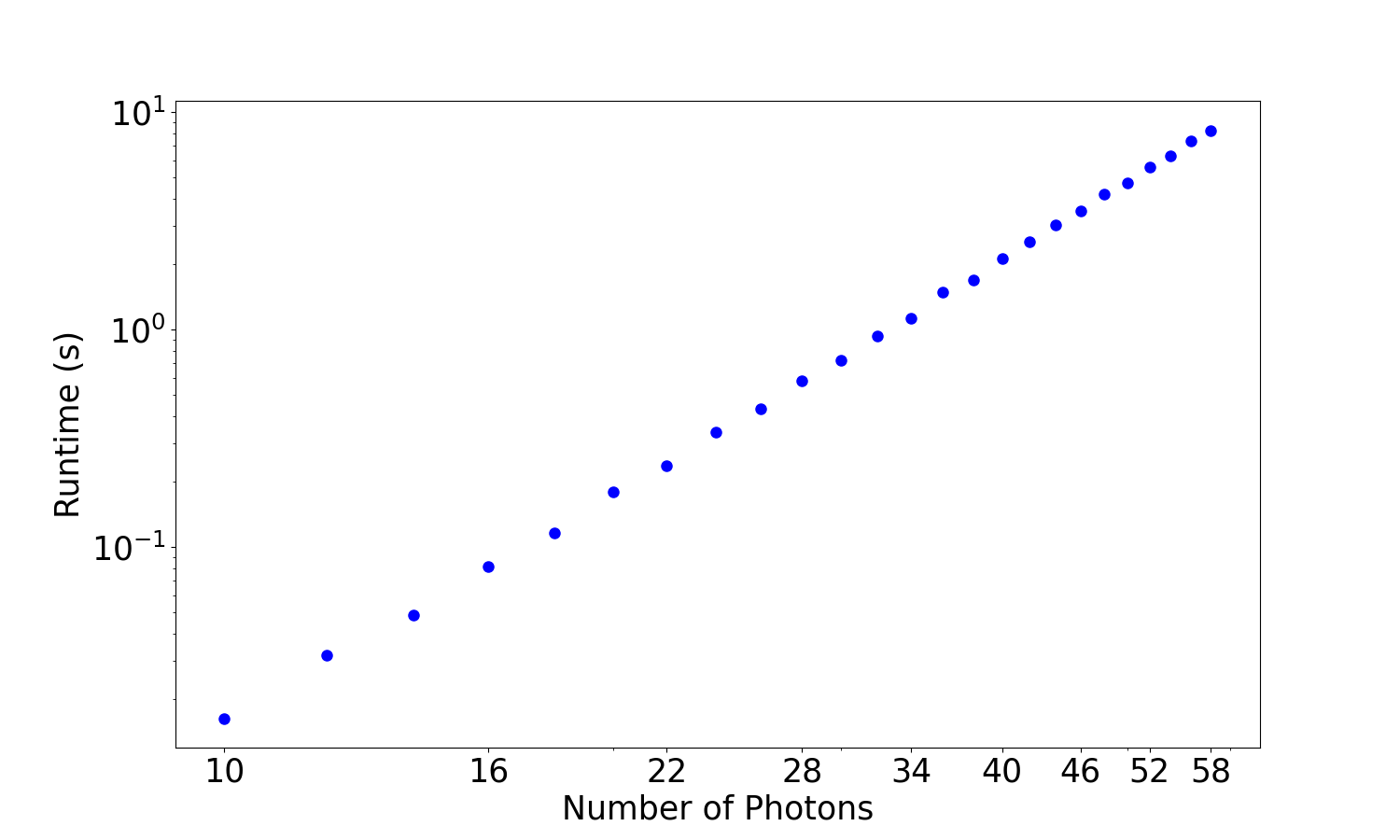}}
	\caption{Runtime of \classname{AlternateTargetSolver} versus the number of photons for (a) linear cluster state (in semi-log scale for y-axis) and (b) repeater graph state (in log-log scale) using an exhaustive search of the graph orbit. A linear line in the semi-log plot indicates the runtime of linear cluster states scales exponentially, whereas a linear line in the log-log plot shows the runtime for repeater graph states scales polynomially. \classname{TimeReversedSolver} is used as the state-to-circuit mapping in the \classname{AlternateTargetSolver}.}
	\label{fig:runtime}
\end{figure}

The runtime of \classname{AlternateTargetSolver} depends on a few choices made by a user. For example, the number of all possible photon emission orderings and the number of local Clifford equivalent graphs can grow exponentially as the number of photons in general. If one attempts to explore every possibility exhaustively, the runtime can scale exponentially. If the user sets a maximum number of alternative graphs for the solver to consider, the runtime will mainly depend on the runtime of the selected state-to-circuit mapping. On the other hand, if a specific family of graphs exhibit certain symmetries, the number of alternative graphs may have a better scaling (e.g. it can grow polynomially with the number of photons). In \Cref{fig:runtime}, we demonstrate the scaling of \classname{AlternateTargetSolver}'s runtime (with \classname{TimeReversedSolver} as the state-to-circuit mapping) for linear cluster states and repeater graph states when an exhaustive search of orbits is employed. In this case, the runtime of linear cluster states exhibits exponential scaling, whereas that of repeater graph states scales polynomially due to the symmetry of repeater graph states. It is also possible to use other methods to search local Clifford equivalent graph states. We refer to \cite{Ghanbari2024} for a correlation-assisted method that avoids the potentially poor scaling of an exhaustive search and achieves performance (in terms of circuit metrics) similar to that of the exhaustive search. 

%% file: text/6conclusion.tex
\section{Conclusion}\label{sec:conclusion}

\software{} is a user-friendly Python framework for optimizing photonic graph state generation circuits. It supports circuit simulation and optimization in the presence of imperfections/noise, flexible cost function definition, and tools for producing several circuits per target output state. Easily extended and customised, \software{} can thus be a valuable tool for researchers studying graph state generation schemes obeying experimentally-relevant constraints. Further information is available in the Appendices as well as \software{}'s documentation.

%% file: text/7back_material.tex
\section*{Code availability}
\software{} is available on \url{https://github.com/graphiq-dev/graphiq} under Apache 2.0 license. The full documentation can be found in \url{https://graphiq.readthedocs.io/en/latest/}. 

\section*{Acknowledgement}
We thank Stefania Sciara for helpful discussions. We thank Kai Sum Chan and Aayush Soni for discussions and their help in code refactoring. This work is supported by MITACS, the National Research Council of Canada under NRC-CSTIP QSP-081-1, the Vanier CGS Program, and the Air Force Office of Scientific Research (AFOSR) under Grant FA9550-22-1-0062. R.G.M. acknowledges financial support from NSERC. H.-K. Lo acknowledges financial support from NSERC, CFI Operating Fund, and Innovative Solutions Canada.

%% file: text/app_background.tex
\section{Background and implementation details}\label{app_sec:background}
In this appendix, we provide a short theoretical background on graph states, the stabilizer formalism, conversion between state representations, noise models, photon emission ordering, and local complementation on graphs,.

\subsection{Graph state}\label{app_subsec:graph}
A graph state can be represented by a graph $G=(V, E)$ consisting of the vertex set $V$ and the edge set $E$. Each vertex (or node) in $V$ represents a qubit in the $\ket{+}:=\frac{1}{\sqrt{2}}(\ket{0}+\ket{1})$ state. Each edge between two vertices represents a Controlled-Z (CZ) gate acting on those two qubits. From a given graph, it is easy to write down the corresponding graph state $\ket{\psi}$ as
\begin{aeq}\label{eq:graph_state_from_graph}
\ket{\psi} = \prod_{(i,j) \in E} CZ_{i,j} \ket{+}^{\otimes |V|},
\end{aeq}where $CZ_{i,j}$ denotes the CZ gate with the control qubit indexed by $i$ and the target qubit indexed by $j$.

% For an $n$-qubit state, we can write the density matrix as
% \begin{aeq}
% \rho = \sum_{i, j=1}^{2^N} \rho_{ij} \ket{i}\bra{j},
% \end{aeq}where $\{\ket{i}: 1 \leq i \leq 2^N\}$ is the set of basis states of the Hilbert space of dimension $2^N$. 
\subsection{Stabilizer formalism}\label{app_subsec:stabilizer}
\subsubsection{Stabilizer group and symplectic representation}
A stabilizer group $S$ on $n$ qubits is a subgroup of the $n$-qubit Pauli group, $P_n$, such that it is Abelian and it does not contain $-I$. An $n$-qubit stabilizer state $\ket{\psi}$ is defined as a simultaneous eigenvector with eigenvalue $1$ of a stabilizer group, $S$. The number of independent generators in the stabilizer group corresponding to an $n$-qubit stabilizer state is equal to $n$.

The binary symplectic representation of each Pauli operator is listed in \Cref{table:binary_symplectic}. The set of $n$ generators for the stabilizer group is then translated to two $n \times n$ matrices, one to represent Pauli $X$ matrix components and the other to represent Pauli $Z$ matrix components. The first digit in the translation becomes an entry in the $X$ component matrix and the second digit becomes an entry in the $Z$ component matrix.
\begin{table}[h]
\centering
\small
    \rowcolors{1}{gray!10}{white}
        \arrayrulecolor{xgreen} 
    \setlength{\tabcolsep}{0pt}
    \setlength\extrarowheight{5pt}
 \begin{tabular}{m{3cm} >{\raggedright\arraybackslash}m{5 cm} }
    \rowcolor{tablehead} Pauli operator & Binary representation \\
     	\toprule
    $I$ & 00 \\
   $X$ & 10  \\
    $Y$ & 11  \\
    $Z$ & 01  \\
     \end{tabular}
\caption{Translation between Pauli operators and their binary symplectic representation.}
\label{table:binary_symplectic}
\end{table}

The stabilizer formalism allows us to efficiently represent so-called stabilizer states, a subset of quantum states, by tracking the generators of the stabilizer group that stabilizes the quantum state \cite{aaronson_improved_2004,audenaert_entanglement_2005}. As \software{} is tailored for photonic graph state generation and graph states form a subset of stabilizer states, it is natural to consider the stabilizer formalism. The number of generators needed to represent a stabilizer state grows linearly with the number of qubits, $n$. Since each generator consists of $n$ Pauli operators, a stabilizer state can be represented by writing down $n^2$ Pauli operators. This stabilizer representation scales well with the number of qubits, making it more efficient for the study of large systems relative to the density matrix formalism. In practice, we adopt a binary symplectic representation to the stabilizer tableau. Specifically, we adopt the formalism in Ref. \cite{aaronson_improved_2004} to improve the efficiency of simulating measurements at a cost of doubling the size of the tableau. We keep track of destabilizer generators in addition to stabilizer generators. Destabilizer generators are Pauli operators such that destabilizer and stabilizer generators together generate the entire $n$-qubit Pauli group (modular four phases). 

\subsubsection{Tableau}
In the \classname{CliffordTableau} class implemented in \software, for an $n$-qubit stabilizer state, the state is represented by an $2n \times 2n$ binary matrix as below:
\begin{equation*}
 \arrayrulecolor{black} 
\left(\begin{array}{@{}c|c@{}}
 \begin{matrix}
  x_{11} & \dots & x_{1n} \\
  \vdots & \ddots & \vdots \\
  x_{n1} & \dots & x_{nn}
  \end{matrix}
  &   \begin{matrix}
  z_{11} & \dots & z_{1n} \\
  \vdots & \ddots & \vdots \\
  z_{n1} & \dots & z_{nn}
  \end{matrix} \\
\hline
   \begin{matrix}
  x_{(n+1)1} & \dots & x_{(n+1)n} \\
  \vdots & \ddots & \vdots \\
  x_{(2n)1} & \dots & x_{(2n)n}
  \end{matrix}& 
  \begin{matrix}
  z_{(n+1)1} & \dots & z_{(n+1)n} \\
  \vdots & \ddots & \vdots \\
  z_{(2n)1} & \dots & z_{(2n)n}
  \end{matrix} 
\end{array}\right)
\end{equation*}where rows $1$ to $n$ represent the destabilizer generators, and rows $n+1$ to $2n$ represent the stabilizer generators. Two additional binary vectors (each of size $2n$) are used represent the phases (1 out of 4 possible values in the set $\{\pm 1, \pm i \}$) of those $2n$ generators:
\begin{equation*}
\begin{array}{cc}
\left(\begin{array}{c}
\begin{matrix} r_1 \\ \vdots \\ r_n\end{matrix}  \\
\hline
\begin{matrix} r_{n+1} \\ \vdots \\ r_{2n} \end{matrix}  
\end{array}\right), 
&
\left(\begin{array}{c}
\begin{matrix} i_1 \\ \vdots \\ i_n\end{matrix}\\
\hline
\begin{matrix} i_{n+1} \\ \vdots \\ i_{2n}\end{matrix} 
\end{array}\right).
\end{array}
\end{equation*}We remark that the second phase vector is needed to properly handle all Clifford gates in the circuit simulation. While the phases for stabilizer generators can be either $1$ or $-1$, the destabilizer generators can take $\pm i$ phases in addition to $\pm 1$ phases. Due to the need to keep track of the phases of destabilizer generators to guarantee the correct simulation of measurement gates, we choose to include this second phase vector. As such, it is different from Ref. \cite{aaronson_improved_2004} in this aspect. 

As an example, the tableau for $\ket{0}^{\otimes n}$ is
\begin{aeq}\label{eq:tableau_0_state}
\left(\begin{array}{c|c}
I_n & 0_n  \\
\hline
0_n & I_n 
\end{array}\right), 
\end{aeq}where $I_n$ is $n \times n$ identity matrix and $0_n$ is the $n \times n$ matrix with all zeros. The two phase vectors of this state are initialized to all zeros.

\subsubsection{Mixed stabilizer}
The stabilizer formalism is suitable for graph state generation since graph states can be generated with only Clifford operations. However, only a subset of quantum operations can be efficiently simulated within the stabilizer representation, limiting the scope for circuit optimization, especially when noise sources are included. This poses the challenge that this representation is unsuited for the simulation of noise without judicious changes to the formalism. We amend this representation with a couple of approaches to handle mixed states. 

A mixed state can be written as an ensemble of pure states. To handle mixed states in the stabilizer representation, one idea is to keep track of all pure states in the ensemble with their associated probability distribution. The \classname{MixedStabilizer} class implements this idea. This representation is suitable when the number of pure states in the ensemble is relatively small. In the simulation, each tableau needs to be updated independently and thus the running time of the simulation also grows with the number of pure states in the ensemble. In practice, if the number of possible events can grow fast when considering a noise model that introduces probabilistic noises (e.g. a depolarizing noise), the memory requirement and the simulation running time for this representation can be very demanding. However, it is not always necessary to keep track of all possible events. For example, if one is interested in a bound on a given metric instead of a precise value, ignoring states with very low probabilities does not loosen the bound significantly while saving substantial computational resources. For this reason, the \classname{MixedStabilizer} class allows the probability distribution to be subnormalized, that is, the sum of all probabilities is smaller or equal to 1.

%\Cref{code:stabilizer} demonstrates an example of creating instances of \classname{CliffordTableau}, \classname{Stabilizer} and \classname{MixedStabilizer} classes.
%\begin{code}
%	\pythonfile{code/stabilizer.py}
%	\caption{An example of creating an instance of \classname{CliffordTableau}, that of \classname{Stabilizer} and that of \classname{MixedStabilizer}.}
 %\label{code:stabilizer}
%\end{code}

\subsection{Conversion between state representations}\label{app_subsec:state_conversion}
Given the clear advantages and disadvantages of various state representations, it is useful to be able to convert one representation to another as necessary. As the set of all graph states is a proper subset of stabilizer states, and the set of all stabilizer states is a proper subset of all quantum states representable by density matrices, it is expected that not all conversions are allowed. Here, we restrict to the cases where a given state is a stabilizer state or a probabilistic mixture of stabilizer states. 

Conversion between representations is mostly straightforward with the exception of converting a density matrix representation to a graph representation. 

From a graph $G=(V,E)$ whose adjacency matrix is $\theta$, the stabilizer generators $R_j$'s corresponding to the graph state are given by
\begin{aeq}\label{eq:stabilizer_from_graph}
R_j = X_{j} \prod_{k=1}^{n} Z_k^{\theta_{kj}},
\end{aeq}%
for $j=1,\dots, n$, where $X_j$ is the Pauli $X$ operator acting on the $j$-th qubit, $Z_k$ is the Pauli $Z$ operator acting on the $k$-th qubit and $\theta_{kj}$ is the $(k,j)$-entry of the adjacency matrix $\theta$.

From the set of generators $\{R_j: j=1,\dots,n\}$, one can construct the density matrix $\rho$ of the stabilizer state by
\begin{aeq}\label{eq:density_from_stabilizer}
\rho = \frac{1}{2^n} \prod_{j=1}^n (I_n + R_j).
\end{aeq}

From a graph $G=(V,E)$, it is straightforward to write down the corresponding density matrix $\rho := \ket{\psi}\bra{\psi}$, where $\psi$ is given in \Cref{eq:graph_state_from_graph}. From the graph, the stabilizer generators can be found by \Cref{eq:graph_state_from_graph}. In our stabilizer representation, we also need to find destabilizer generators. \software{} includes a function to find gates that can disentangle an $n$-qubit stabilizer state into $\ket{0}^{\otimes n}$. As the \classname{CliffordTableau} for $\ket{0}^{\otimes n}$ is given in \Cref{eq:tableau_0_state}, one can construct the \classname{CliffordTableau} for the stabilizer state by applying those gates in the reversed order. The conversion from a stabilizer formalism to a density matrix simply follows \Cref{eq:density_from_stabilizer}.

A less straightforward conversion is from a density matrix representation to a graph representation. For this conversion, we have developed a procedure that constructs an adjacency matrix from a density matrix. To construct the graph, we must verify the existence of an edge between each pair of qubits in the state. To do so, we apply Pauli $Z$ measurements to all qubits except those indexed by $i$ and $j$. All these measurements effectively disentangle all other qubits and keep these two qubits indexed by $i$ and $j$. After obtaining the reduced density matrix of these two remaining qubits, we calculate the negativity of the reduced density matrix \cite{plenio2007introduction,schwartz_deterministic_2016} to verify if an edge exists between these two qubits. Since the space is reduced to two qubits, the negativity serves as an accurate entanglement metric and gives a positive value when two nodes are connected \cite{Peres_separability1996,Horodecki_separability1996}. By applying this procedure to every pair of nodes, we can construct the adjacency matrix from the density matrix. We note that this procedure works under the assumption that our state is a graph state.  

\begin{breakablealgorithm}
 \label[algorithm]{alg:density_to_graph}
 \caption{Converting the density matrix of a graph state to its adjacency matrix graph representation.}
 \textbf{Inputs:} \\[0.1cm]
	\begin{tabular}{ll}
		$\rho$ & A density matrix that corresponds \\
    & to an $n$-qubit graph state \\
    $\delta \in [0.49, 0.5]$ & A threshold for negativity
	\end{tabular}
	\hspace{0.3cm}
	
 \textbf{Output:} \\[0.1cm]
	\hspace{0.3cm}
	\begin{tabular}{ll}
		$\theta$ & An adjacency matrix
	\end{tabular}
 	\vspace{0.3cm}
  
	\textbf{Algorithm:}\\
 \begin{enumerate}	
 \addtocounter{enumi}{0}
 \item Initialize $\theta := 0_n$; \\
  \item for $i = 1$ to $n-1$: \\
    \; \; for $j = i+1$ to $n$:
	\begin{enumerate}
		\itemsep0em
		\item Obtain $\rho^{(i,j)}$ by measuring all qubits, except qubits indexed by $i, j$, in the Z basis;
		\item Calculate negativity of the two qubit system $\rho^{(i,j)}$;
            \item If  $\rho^{(i,j)} \geq \delta$, set $\theta(i, j) = \theta(j, i) = 1$.
        \end{enumerate}
   \item  Return $\theta$.
    \end{enumerate}
    
\end{breakablealgorithm}

We note that one approach to translate a density matrix $\rho=\ket{\psi}\bra{\psi}$ corresponding to a pure stabilizer state to a stabilizer representation is as follows. First, we find the stabilizer group corresponding to the stabilizer state $\ket{\psi}$ by iterating through each element $M$ in the $n$-qubit Pauli group and recording $M$ if it stabilizes the state $\ket{\psi}$, that is, $M\ket{\psi} = \ket{\psi}$. After obtaining the stabilizer group, we can find a set of generators that generate the whole group. Then, using that set of generators, we can write down the binary symplectic representation. While this procedure can return a correct answer, it is very inefficient due to the size of the $n$-qubit Pauli group. As we have an efficient ($O(n^2)$) approach to convert a density matrix to the graph representation and also an efficient way to convert a graph representation to the stabilizer representation, we can combine these two methods to move from the density matrix to the stabilizer representation for graph stabilizer states.

\subsection{Noise models}\label{app_subsec:noise_models}
In \software{}, we allow for the flexible placement of noise. A noise source can arise before or after a quantum gate. It can also replace a gate with another gate. The noise models presently implemented in \software{} focus on common noise types for photonic graph state generation, particularly, depolarizing noise and photon loss. We also include additional noise models that can be useful for versatile circuit simulation purposes.

As discussed before, each state representation has a varying capacity for representing mixed states. Since noise sources often turn pure states into mixed states, each state representation will have a different ability to handle noise. Both the density matrix and stabilizer backends support depolarizing noise and photon loss as well as Pauli errors. The density matrix backend also supports arbitrary unitary or mixed unitary errors.

%Noise models implemented in \software{} are shown in \Cref{table:noise_models}. 
%\begin{table}[!ht]
%\centering
%\small
%    \rowcolors{1}{gray!10}{white}
%        \arrayrulecolor{xgreen} 
%    \setlength{\tabcolsep}{0pt}
%    \setlength\extrarowheight{5pt}
% \begin{tabular}{m{3.7cm} >{\raggedright\arraybackslash}m{2cm}  >{\raggedright\arraybackslash}m{2cm}}
%    \rowcolor{tablehead} Noise model & Density matrix & Stabilizer/ Mixed stabilizer/ Graph\\
%     	\toprule
%    Depolarizing noise & \checkmark & \checkmark \\
 
%    Photon loss & \checkmark & \checkmark\\
 
%    Pauli error & \checkmark & \checkmark\\
 
%    Arbitrary unitary error & \checkmark & \xmark\\
%     \end{tabular}

%\caption{\label{table:noise_models}Noise model implementations in \software{} for different state representation.}
%\end{table}
We discuss depolarizing noise and photon loss below.

\subsubsection{Depolarizing noise}
Depolarizing noise is often used to model qubit errors. A depolarizing channel acting on a density matrix $\rho$ is defined as
\begin{aeq}
\mathcal{D}_p(\rho) = (1-p) \rho + \frac{p}{3}(X \rho X + Y \rho Y + Z \rho Z),
\end{aeq}where $X, Y, Z$ are Pauli matrices, and $p$ is the depolarizing probability. The resulting state $\mathcal{D}_p(\rho)$ is in general a mixed state. The density matrix is the most versatile as it can inherently and easily include any type of noise. However, it soon becomes impractical when the number of qubits increases. The stabilizer formalism works well for pure states but cannot directly represent mixed states. To handle noises like depolarizing noise in the stabilizer representation, it is necessary to extend the representation, as in the mixed stabilizer representation implemented in the \classname{MixedStabilizer} class. Alternatively, one may also consider Monte Carlo simulation to simulate noisy events. \software{} has an implementation of Monte Carlo simulation for this purpose. Another approach is to keep track of the most probable event(s) using the \classname{MixedStabilizer} class.

\subsubsection{Photon loss}
When a photon is lost, the corresponding qubit moves outside the computational space. One approach to represent photon loss is to use a qutrit description where the first two dimensions are the original qubit space and the third dimension represents the vacuum state. This extension naturally allows the density matrix representation to handle loss. Another approach is to calculate the probability of the event where no photon is lost. Arguably, this event is the event of interest for photonic graph state generation. Metrics like fidelity dismiss all other events since the target state has zero overlap with any state that loses one or more photons. In \software{}, we thus implement the latter idea. 

We remark that it is possible to make a protocol loss-tolerant by applying some quantum error correction code to protect logical qubits \cite{varnava_loss_2006}. The present version of \software{} (version 0.1.0) is restricted to the case where no such encoding is performed. However, we expect to implement an error correction module in future versions of \software{}.

\subsection{Photon emission ordering}\label{app_subsec:emission_ordering}
As multiple photons in a graph state can be emitted by one common quantum emitter, it is important to consider in what order these photons should be emitted. Different emission orderings are likely to lead to different quantum circuits since the local entanglement structures might be quite different for different photons. Consequently, the user needs to specify the photon emission ordering in the target graph state, which we use to number the graph nodes, i.e, we label nodes numerically where a smaller number means an earlier time slot in the emission process. 

As discussed in Ref. \cite{li_photonic_2022}, different emission orderings require different minimal numbers of quantum emitters to generate the graph state deterministically. However, the task of finding optimal photon emission orderings for an arbitrary graph is an NP-hard problem \cite{li_photonic_2022}. Nevertheless, \software{} provides the users with the option to automatically consider different permutations of the emission ordering while solving for a generating circuit. This search can be either exhaustive or driven by the random sampling of all possible $n!$ permutations. Note, however, that not every permutation yields a distinct new graph since a subset of permutations result in graph automorphisms. \software{}'s relabelling module removes the automorphic cases so that the returned set of graphs are all distinguishable. A user can specify parameters to choose how many different emission orderings to consider and how to generate a selective subset of orderings.
%For graphs with certain symmetries, it might be possible to find the optimal ordering in polynomial time since the possible number of orderings (up to graph automorphism) can in principle scale slower than $O(n!)$. In \software{}, our goal is not to find an efficient heuristic algorithm that solves this NP-hard problem for a large subset of graph states. Instead, we use different photon emission orderings to explore alternative quantum circuits. For this purpose, a user can specify parameters to choose how many different emission orderings to consider and how to generate a selective subset of orderings.

\subsection{Local Clifford and local complementation}\label{app_subsec:lc}
%Local Clifford gates are tensor products of one-qubit Clifford gates. As Clifford gates are efficiently simulatable by classical computers, local Clifford gates are typically easy to apply in an experimental setup. Interestingly, applying local Clifford gates to a graph state is closely related to local complementation actions on the corresponding graph.

\begin{figure}[t]
\centering
\subfloat[Initial graph]{\label{fig:lc_before}\includegraphics[width=0.35\linewidth]{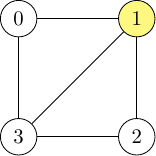}}
\hspace{1cm}
\subfloat[Apply local complementation on node 1]{\label{fig:lc_after}\includegraphics[width=0.35\linewidth]{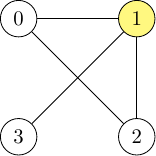}}
	\caption{Illustration of local complementation. By applying the local complementation operation on  node 1, neighbors of node 1 that were connected to each other in the initial graph are now disconnected and those that were disconnected are now connected.}
\label{fig:local_complementation}
\end{figure}

For a graph $G=(V,E)$, a local complementation on a node $v \in V$ corresponds to applying the complementation to a subgraph consisting of all neighbors of $v$. Let $N(v)$ denote all the neighboring nodes of $v$ (i.e., $w \in V$ is in $N(v)$ if and only if $(v,w) \in E$.) Applying the local complementation on the node $v$ generates another graph $G' = (V, E')$ where for each pair of $u, w \in N(v),$ if $(u, w) \in E$, then $(u, w) \not\in E'$; if $(u, w) \not\in E$, then $(u, w) \in E'$; all other edges in $E$ are also in $E'$.  \Cref{fig:local_complementation} shows an example of applying local complementation on node $1$.

The relation between local complementation actions on a graph and local Clifford gates applied to the corresponding graph state is discussed in Ref. \cite{van_den_nest_graphical_2004}. Each local complementation action corresponds to a sequence of local Clifford gates. If a graph can be transformed into another graph via local complementation actions, then the corresponding graph states are local Clifford equivalent. An efficient algorithm to find local Clifford operations between two local Clifford equivalent states is given in Ref. \cite{van_den_nest_efficient_2004}. The overall complexity of this algorithm is $O(n^4)$.

%% file: text/app_examples.tex
\onecolumn
\section{Examples}\label{app_sec:examples}

\subsection{Code examples}\label{app_sec:code_blocks}

\Cref{code:SimulationCodeBlock} demonstrates an example of simulating a quantum circuit that generates a photonic 3-qubit linear cluster using the density matrix backend.

\begin{code}
	\pythonfile{code/simulate.py}
	\caption{Simulating a quantum circuit that produces the three-qubit linear cluster state in \Cref{fig:graph_linear3} using the density matrix backend.}
 \label{code:SimulationCodeBlock}
\end{code}

\Cref{code:evaluation} presents an example of evaluating the quantum circuit found by \classname{TimeReversedSolver} using the infidelity and the circuit depth metrics.

\begin{code}
	\pythonfile{code/evaluate.py}
	\caption{Using the \classname{TimeReversedSolver} to find a quantum circuit that generates the 3-qubit linear cluster state, followed by evaluating the output state's infidelity and circuit depth.}
 \label{code:evaluation}
\end{code}

\Cref{code:exploration} presents an example of using \classname{AlternateTargetSolver}. This example uses the stabilizer backend for the noise-free simulation and uses the density matrix backend for the noise simulation. The noise model used is depolarizing channels. More examples can be found in the Jupyter notebooks in the \software{}'s code repository. 

\begin{code}
	\pythonfile{code/exploration.py}
	\caption{\label{code:exploration}Using the \classname{AlternateTargetSolver} to find quantum circuits that generate a target graph state.}
 
\end{code}

\FloatBarrier

\subsection{Sample results using \software{}}\label{app_subsec:alternative_circuits}
\subsubsection{3-qubit linear cluster state}\label{app_subsec:linear_cluster}
\begin{figure}[h]
\subfloat[Using \classname{TimeReversedSolver}]{\label{fig:example_linear3_deterministic}\includegraphics[width=0.5\linewidth]{fig/fig3b.png}} 
\subfloat[Using \classname{EvolutionarySolver}]{\label{fig:example_linear3_evolutionary}\includegraphics[width=0.5\linewidth]{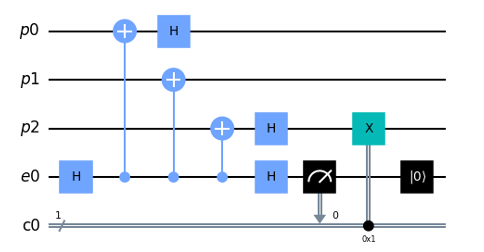}}
	\caption{Optimal circuits for generating the three-qubit linear cluster state when the cost function is the infidelity function, using (a) the \classname{TimeReversedSolver} based on Ref. \cite{li_photonic_2022} and (b) using \classname{EvolutionarySolver} with the random search option.}
\label{fig:example_linear3}
\end{figure}

\Cref{fig:example_linear3} shows an example of finding alternative quantum circuits for the three-qubit linear cluster state using our probabilistic solvers, compared against the output of the deterministic solver based on Ref. \cite{li_photonic_2022}. We see that our probabilistic solvers can potentially produce quantum circuits with reduced circuit depth. 

\subsubsection{Improving fidelity}\label{app_subsec:example_alternate_graph_0}

\begin{figure}[h]
\centering
\subfloat[original]{\label{fig:graph_exp1}
\includegraphics[width=0.25\linewidth]{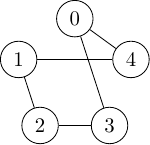}}
\subfloat[alternative]{\label{fig:graph_exp2}
\hspace{0.25cm}\includegraphics[width=0.25\linewidth]{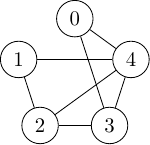}}
	\caption{Original target graph and its alternative, locally equivalent shape found using the \classname{AlternateTargetSolver}.}
\label{fig:graph_exp}
\end{figure}

Using the graph shown in \Cref{fig:graph_exp1} as the target graph for \classname{AlternateTargetSolver}, we search for a generation circuit that offers a better fidelity for the final state with respect to when the circuit is obtained using the \classname{TimeReversedSolver}.
\begin{figure}[H]
	\centering
	\scalebox{.57}{
		\includegraphics{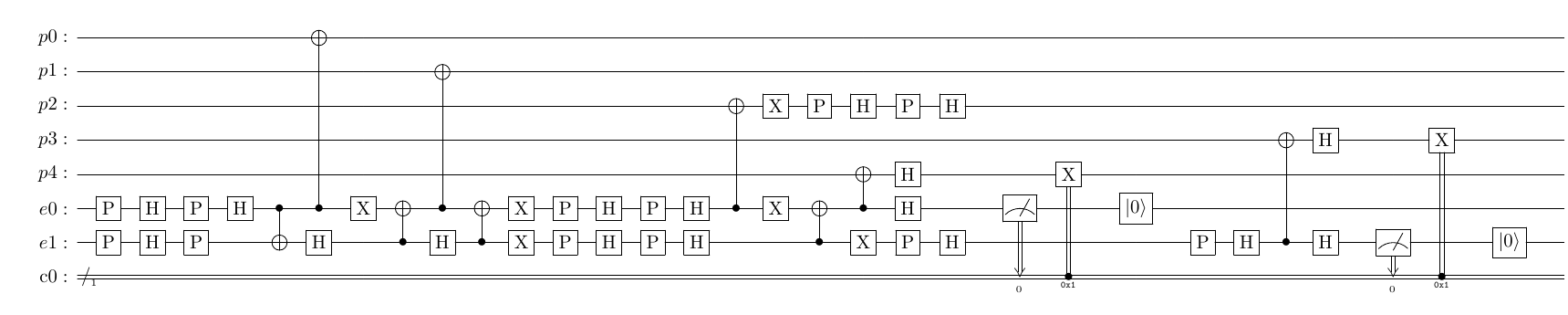}
	}
	\scalebox{.57}{
		\hspace{-10cm}\includegraphics{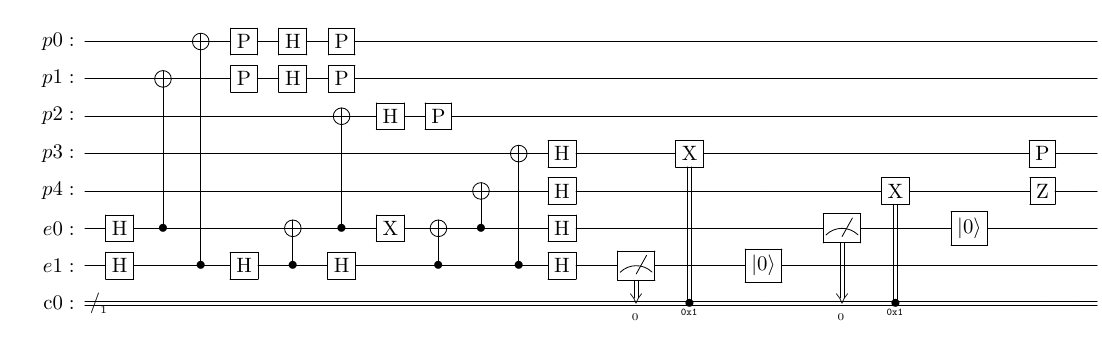}
	}
	\caption{The original (top) and alternative (bottom) circuits, corresponding to the use of \classname{TimeReversedSolver} and \classname{AlternateTargetSolver} respectively.}
	
	\label{fig:circuit_improvement}
\end{figure}

The circuits for the original graph (found via \classname{TimeReversedSolver}) and the alternative graph (found via \classname{AlternateTargetSolver}) are depicted in \Cref{fig:circuit_improvement}. When applying a depolarizing noise model on the emitter gates with a 0.01 depolarization rate, the fidelity is increased from 0.66 to 0.81 for the alternative case. The reduction in the size of the circuit is also evident in \Cref{fig:circuit_improvement}.

\subsubsection{3D cluster state $2\times2\times3$}\label{app_subsec:example_alternate_graph}

\begin{figure}[h]
    \centering
    \scalebox{2}{\includegraphics{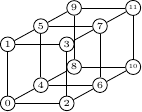}}
    \caption{A 12-qubit $2\times2\times3$ 3D cluster state.}
    \label{fig:3dcluster}
\end{figure}

This example demonstrates the application of \classname{AlternateTargetSolver}
to explore different generation circuits using the local Clifford equivalency orbit of a 3D cluster state (\Cref{fig:3dcluster}). 
\Cref{tab:3dcluster} shows the range of values for a set of circuit metrics, encompassing the total number of unitary gates, the number of CNOT gates between emitters, and the circuit depth, across all available options. 
\begin{table}[]
    \centering
        \begin{tabularx}{\columnwidth}{|X|X|X|X|}
        \hline
        Metrics                          & Unitaries & CNOTs & Depth \\ \hline
        Best case                        & 46              & 12          & 16            \\ \hline
      Worst case & 203             & 28          & 59            \\ \hline
        \end{tabularx}
    \caption{The best and worst values for a set of circuit metrics over 5160 members of the LC equivalency class of a three-dimensional $\left(2\times2\times3\right)$ cluster state. The metrics correspond to the number of unitary gates, number of emitter-emitter CNOT gates, and the depth of the corresponding quantum circuits.}
    \label{tab:3dcluster}
\end{table}